# Electrically Tunable Macroscopic Quantum Tunneling in a Graphene-based Josephson Junction

Gil-Ho Lee,[1] Dongchan Jeong,[1] Jae-Hyun Choi,[1, *] Yong-Joo Doh,[2, †] and Hu-Jong Lee[1, †]

[1]*Department of Physics, Pohang University of Science and Technology, Pohang 790-784, Republic of Korea*
[2]*Department of Display and Semiconductor Physics,
Korea University Sejong Campus, Chungnam 339-700, Republic of Korea*
(Dated: September 1, 2011)

Stochastic switching-current distribution in a graphene-based Josephson junction exhibits a crossover from the classical to quantum regime, revealing the macroscopic quantum tunneling (MQT) of a Josephson phase particle at low temperatures. Microwave spectroscopy measurements indicate a multi-photon absorption process occurring via discrete energy levels in washboard potential well. The crossover temperature for MQT and the quantized level spacing are controlled with the gate voltage, implying its potential application to gate-tunable superconducting quantum bits.

PACS numbers: 85.25.Cp, 74.45.+c, 72.80.Vp

A Josephson junction [1], consisting of two superconducting electrodes connected to each other via a nonsuperconducting spacer, provides the basic building block of superconducting quantum bits (qubits) [2] for constructing scalable solid-state quantum computers [3]. Recent advances in nanofabrication techniques have enabled the fabrication of nanostructured proximity-coupled Josephson junctions based on conducting spacers such as nanowires [4, 5], carbon nanotubes [6, 7], nanocrystals [8], and graphene [9]. In contrast to typical tunneling or weak-link [10] Josephson junctions, the Josephson coupling energy in nanostructured Josephson junctions can be tuned electrically for applications in quantum supercurrent transistors [4–6].

In this letter, we report the stochastic switching-current distribution in a superconductor-graphene-superconductor (SGS) junction, demonstrating macroscopic quantum tunneling (MQT) [11, 12] behavior at low temperatures and energy level quantization [12] in a potential well for a Josephson phase particle. The crossover temperature for MQT and the energy level separation are controlled by applying a gate voltage to the SGS junction. Observation of this gate-tunability of the quantum phenomena, first in proximity-coupled Josephson junctions, is mainly indebted to the gate-tunable Josephson coupling energy in the SGS junction. Our findings are relevant to enhanced degrees of freedom for manipulating the macroscopic quantum state in a Josephson junction and to the development of a new type of quantum devices such as a gate-tunable superconducting phase qubit.

SGS hybrid devices were assembled from single-layer graphene in contact with superconducting $Pb_{0.93}In_{0.07}$ (PbIn) electrodes. We alloyed Pb with In to minimize the granularity of electrodes while keeping $T_c$ almost intact [13, 14]. Figure 1(a) shows a representative optical microscopy image of the PbIn-graphene-PbIn Josephson junction. PbIn superconducting electrodes significantly enhances the critical current ($I_c$) compared with com-monly used Al (as high as $I_c \sim 6.0$ μA in highly doped regions) for both the electron and hole side at the base temperature ($T = 50$ mK). This leads to a much stronger Josephson coupling energy, $E_J = \hbar I_c/2e$, where $\hbar$ is Planck's constant ($h$) divided by $2\pi$. The normal-state resistance ($R_N$) of the SGS Josephson junction exhibits a charge-neutral Dirac point occurring at the back-gate voltage of $V_{DP} = -2.5$ V [inset of Fig. 1(b)]. The corresponding mean free path and diffusion constant of the graphene layer are estimated to be $l_m^* \sim 22$ nm and $D \sim 110$ cm$^2$/s, respectively [14]. Because $l_m^*$ and the Thouless energy ($E_{Th} = \hbar D/L^2 \sim 80$ μeV) are much smaller than the junction spacing of $L = 300$ nm and the superconducting energy gap of $\Delta_{PbIn}=1.1$ meV, respectively, our SGS junctions are subject to a long diffusive junction limit [15].

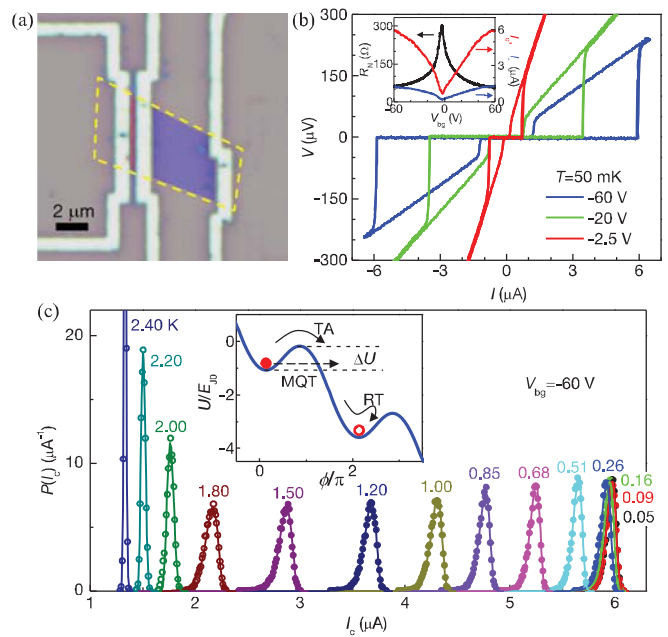

FIG. 1. (color online) (a) Optical image of the device. A graphene flake was placed in the region bound by the yellow dotted line, and the junction area is denoted in red. (b) Typical $I-V$ curves for various $V_{bg}$ at $T = 50$ mK. Inset: $V_{bg}$ dependence of $R_N$ (black), $I_c$ (red), and $I_r$ (blue). $R_N$ was measured in a perpendicular magnetic field of $H = 50$ mT. (c) $I_c$ distribution obtained at $V_{bg} = -60$ V. Solid lines are best fits to the thermal activation (TA, solid symbols) and phase diffusion (PD, void symbols) models. Inset: Schematic of washboard potential and dynamics of a phase particle escaping from the local minimum.

The current-voltage ($I-V$) curves for different $V_{bg}$ in the superconducting state at the base temperature [Fig. 1(b)] exhibit hysteresis for the switching currents of $I_c$ (from the supercurrent branch to the resistive branch) and $I_r$ (vice versa). Hysteretic behavior is common in SGS junctions [9]. It also has been observed in other nanostructured Josephson junctions [4–7], where it was attributed to self-heating [16] or an effective capacitance ($C_{eff} = \hbar/R_N E_{Th}$) due to diffusive motion of quasi-particles in graphene [17]. Our numerical calculations based on the resistively and capacitively shunted junction (RCSJ) model [18] fit the data well with a finite junction capacitance $C$ [14]. Thus, the switching-current distribution $P(I_c)$ in our SGS junction [14] is explicable in terms of underdamped-junction motion of a fictitious phase particle in a tilted washboard potential [18] [inset of Fig. 1(c)]. Here, the mass of the phase particle is proportional to $C$ and it is assumed to move in a metastable potential well of $U(\phi) = -E_{J0}[\cos(\phi) + (I/I_{c0})]$ where $\phi$ is the phase difference across the junction, $I_{c0}$ and $E_{J0}(= \hbar I_{c0}/2e)$ are the fluctuation-free switching current and Josephson coupling energy, respectively. The escape rate of a phase particle from the potential well corresponds to the switching rate from the supercurrent state to the resistive state, which is governed by the Josephson coupling energy and thermal and quantum fluctuations.

$P(I_c)$ in our SGS junction is shown in Fig. 1(c) for $T = 0.05 \sim 2.40$ K at $V_{bg} = -60$ V, where three distinct temperature regions are identified. For $T > 1.80$ K, a symmetric, narrow $P(I_c)$ is obtained, which broadens with decreasing $T$. For $0.51$ K $< T < 1.80$ K, the distribution is asymmetric, and the width narrows with decreasing $T$; this $T$ dependence is opposite to that for $T > 1.80$ K. For $T < 0.26$ K, the shape and width of $P(I_c)$ are nearly independent of $T$. Theoretical fits lead to the conclusion that each of the three temperature regions corresponds to the phase diffusion (PD) [19, 20], thermal activation (TA) [21], and macroscopic quantum tunneling (MQT) [12] of a phase particle, respectively. In this letter, we focus on the TA and MQT regimes, and the PD regime is discussed in supplemental material [14].

In the TA regime, switching from the superconducting to the resistive state occurs via thermally activated escape of the phase particle from a local minimum of a tilted-washboard potential. For quantitative analysis, the switching probability of the relationship $P(I_c) = [\Gamma(I_c)/(dI/dt)]\{1 - \int_0^{I_c} P(I')dI'\}$ is used [21], where $dI/dt$ is the current-sweeping rate and $\Gamma$ is the escape rate. The rate of thermal escape is given by $\Gamma_{TA} = a_t(\omega_p/2\pi) \exp[-\Delta U/k_B T]$, where $a_t = (1 + 1/4Q^2)^{1/2} - 1/2Q$ is a damping-dependent factor, $\omega_p = \omega_{p0}(1-\gamma^2)^{1/4}$, and $\Delta U = 2E_{J0}[(1 - \gamma^2)^{1/2} - \gamma \cos^{-1} \gamma]$, with the quality factor $Q = 4I_c/\pi I_r$, a Josephson plasma frequency $\omega_{p0} = (2eI_{c0}/\hbar C)^{1/2}$, and $\gamma = I/I_{c0}$. It is noted that the normalized standard deviation (SD) of $P(I_c)$ decreases with decreasing $T$ [Fig. 2(a)].

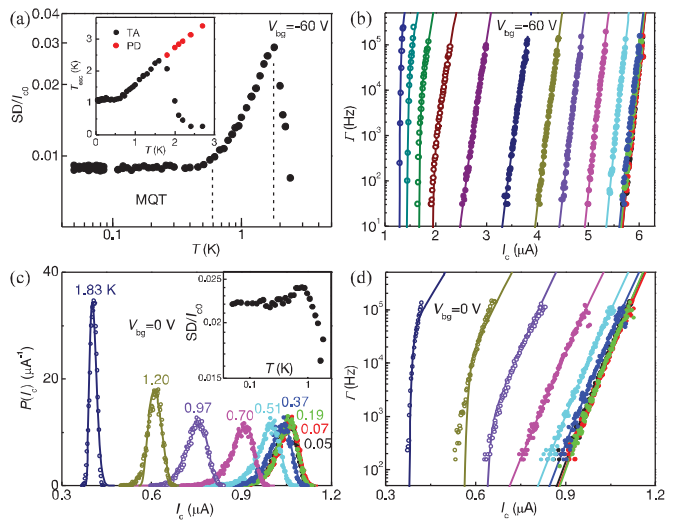

FIG. 2. (color online) (a) $T$ dependence of the normalized standard deviation (SD) of the $I_c$ distribution, $P(I_c)$, obtained for $V_{bg} = -60$ V. Inset: $T_{esc}$ versus $T$ plot obtained from a fit to the TA (black) and PD (red) models. The error bars are smaller than the symbol size. (b) Escape rate $\Gamma$ (symbols) as a function of bias current at $V_{bg} = -60$ V. Solid lines represent fits to the TA (filled symbols) and PD (void symbols) models. Each data set corresponds to the one in Fig. 1(c) with the same color code. (c) $I_c$ distribution obtained for varying $T$ at $V_{bg} = 0$ V. The right-most three curves are not distinguishable. Inset: $T$ dependence of the normalized standard deviation. (d) Escape rate (symbols) overlaid with best-fit curves to the TA (filled symbols) and PD (void symbols) models.

In the MQT regime, however, the escape of the phase particle is governed by the tunneling process with a rate [12] of $\Gamma_{MQT} = 12\omega_p(3\Delta U/\hbar\omega_p)^{1/2} \exp[-7.2(1 + 0.87/Q)\Delta U/\hbar\omega_p]$. As $\Gamma_{MQT}$ is independent of $T$, the normalized SD of $P(I_c)$ in the MQT regime does not vary with $T$ [Fig. 2(a)]. Thus, the $P(I_c)$ and $\Gamma(I_c)$ curves for different $T$ are essentially identical in the MQT regime [see Figs. 1(c) and 2(b)]. The best fit of the escape-rate data to the expression of $\Gamma_{MQT}$ results in $I_{c0} = 7.00$ μA and a junction capacitance of $C = 11.5$ fF. More details of phase-particle escape fitting are given in supplemental material [14].

Fitting the $P(I_c)$ data to the TA model over the entire temperature range of our study reveals the escape temperature ($T_{esc}$), i.e., the temperature perceived by the escaping phase particle. A direct comparison between $T_{esc}$ and $T$ [inset of Fig. 2(a)] shows saturation of $T_{esc}$ in the MQT regime and an almost linear increase of $T_{esc}$ with $T$ in the TA regime. The saturation of $T_{esc}$ corresponds to the temperature region where the TA model fails. The crossover temperature between the two regimes is consistent with that obtained from the normalized SD of $P(I_c)$. With increasing $T$, a sudden decrease in $T_{esc}$ occurs at $T = 1.8$ K, again implying a failure of the TA model at high $T$. A new fit based on the PD model remedies the discrepancy. $T_{esc}$ is higher than $T$ by $\sim 0.5$ K in the TA regime, implying that $\Delta U$ is slightly overestimated in the fit. A non-sinusoidal current-phase relationship [22] exhibited in the SGS junction most likely causes the deviation of $T_{esc}$ from $T$.

Figure 2(c) shows the $T$ dependence of $P(I_c)$ curves at $V_{bg} = 0$ V near $V_{DP}$, displaying a gradual change in the distribution, the overall shape of which is similar to those obtained from the highly doped region at $V_{bg} = -60$ V. The normalized SD of $P(I_c)$ in the inset exhibits broadening (PD regime), narrowing (TA regime), and saturation (MQT regime) behaviors with decreasing $T$. Our data fit well with each model in the corresponding $T$ range [see Figs. 2(c) and 2(d)]. Despite the overall similarity between the two data sets of $P(I_c)$ with different $V_{bg}$, crossover temperatures between the two escape processes (MQT-TA or TA-PD) are suppressed as $V_{bg}$ approaches $V_{DP}$ (see below). The theoretical fit results in values of $I_{c0}$ (=1.91 $\mu$A) and $C$ (=2.4 fF) that are lower than those obtained at $V_{bg} = -60$ V.

Figure 3(a) shows the progressive change in the normalized SD versus $T$ curve with varying $V_{bg}$. The crossover temperature, $T^*_{MQT}$ ($T^*_{TA}$) between the MQT (TA) and TA (PD) regimes, can be defined as the crossing point of two lines extrapolated from each regime. As the applied $V_{bg}$ approaches $V_{DP}$, $T^*_{MQT}$ and $T^*_{TA}$ decrease and the crossover regions broaden gradually.

The crossover temperatures that separate the three distinct regions of PD, TA, and MQT vary with $V_{bg}$ [Fig. 3(b)]. Both $T^*_{MQT}$ and $T^*_{TA}$ decrease sharply near $V_{DP}$, indicating that the escape of a phase particle from the washboard potential can be tuned by applying $V_{bg}$. This constitutes the central observation of this study, the physical origin of which stems from the gate-tunability of $E_{J0}$ by $V_{bg}$ in an SGS junction. Theoretically, $T^*_{TA}$ is linearly proportional to $E_{J0}$ via the relationship [23] $T^*_{TA} \sim E_{J0}[1 - (4/\pi Q)]^{3/2}/30 k_B$. Because $Q$ varies smoothly as a function of $V_{bg}$ between 5 (near $V_{DP}$) and 6 (for $V_{bg} \approx \pm 60$ V), the ratio of $k_B T^*_{TA}/E_{J0}$ is nearly independent of $V_{bg}$ [Fig. 3(d)]. Thus, $T^*_{TA}$ is determined almost solely by the $V_{bg}$ dependence of $E_{J0}$. $T^*_{MQT}$ is related to $\omega_p$ as [24] $T^*_{MQT} = a_t \hbar \omega_p/2\pi k_B$. Because $\omega_{p0}$ is proportional to $I_{c0}^{1/2}$, the resulting re-lationship $T^*_{MQT} \propto I_{c0}^{1/2}$ [Fig. 3(c)] gives a qualitative explanation for the $V_{bg}$ dependence of $T^*_{MQT}$. The junction capacitance, $C^* = 35.2$ fF, estimated from $T^*_{MQT}$ at $V_{bg} = -60$ V is comparable to the result obtained from the $\Gamma_{MQT}$ fit.

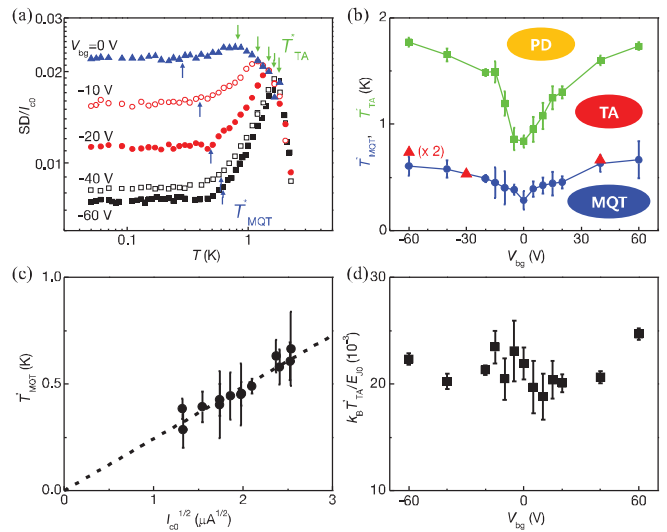

FIG. 3. (color online) (a) Normalized SD versus $T$ plot with $V_{bg} = -60$, -40, -20, -10, and 0 V from bottom to top. Arrows indicate $T^*_{MQT}$ and $T^*_{TA}$ for each value of $V_{bg}$. (b) $V_{bg}$ dependence of the crossover temperatures of $T^*_{MQT}$ (blue circle) and $T^*_{TA}$ (green square). $T^*_{MQT}$ estimated from microwave spectroscopy measurements are also shown, multiplied by a factor of 2 (red triangle). (c) Proportionality relationship between $T^*_{MQT}$ and $I_{c0}^{1/2}$. The dotted line is a guide for the eyes. (d) $T^*_{TA}$ normalized by $E_{J0}$ as a function of $V_{bg}$.

In the MQT regime, the energy levels in the potential well are presumed to be quantized with a level spacing of $E_{10}$ between the ground and the first excited states [12]. For irradiation with microwaves of frequency $f_{mw}$ corresponding to the photon energy equal to the quantized level spacing ($E_{10} = h f_{mw}$), tunneling of a phase particle is possible via the first excited state. In our SGS junction, a slight increase of the microwave power causes a shift of $P(I_c)$ to a lower current, intermediated by a double-peaked distribution [Figs. 4(a) and (b)], exhibiting a resonant peak at the current $I_{res}$. The occurrence of the resonant peak in $P(I_c)$ for $T < T^*_{MQT}$ is attributed to the MQT process via the first excited state in the potential well, the tunneling rate of which is exponentially enhanced above the direct-tunneling rate from the ground state [12]. The corresponding $\Gamma(I)$ curve in Fig. 4(c) shows the enhanced tunneling rate at lower current, signifying resonant transitions between the two energy levels.

In a parabolic approximation of the potential well, the energy-level separation depends on the bias current: $E_{10} = \hbar \omega_{p0}[1 - (I/I_{c0})^2]^{1/4}$. Because $\nu_{p0}$ (= $\omega_{p0}/2\pi$) was substantially higher than $f_{mw}$(=2-20 GHz) used in this measurement, a multi-photon absorption process [25, 26] ($E_{10} = nhf_{mw}$, where $n$ is the number of photons) was preferred to a single-photon process, resulting in the resonant-transition condition of $f_{mw} = (\nu_{p0}/n)[1 - (I_{res}/I_{c0})^2]^{1/4}$. Figure 4(d) displays the observed $I_{res}$ with varying $f_{mw}$ at $V_{bg} = -60$ V, which fits the resonant-condition curves with $\nu_{p0} = 71.7 \pm 0.5$ GHz and $I_{c0} = 8.05 \pm 0.26$ $\mu$A for the number of photons ($n = 4 - 10$) in our measurement range. As $\nu_{p0}$ depends on the applied $V_{bg}$ [14] the quantized energy-level separation in an SGS junction can be tuned not only by the bias current but also by $V_{bg}$. From the $V_{bg}$-dependent $\nu_{p0}$, one can also estimate $T^*_{MQT}$ for comparison with $T^*_{MQT}$ from $P(I_c)$ [see Fig. 3(b)]. Despite the discrepancy between the two different $T^*_{MQT}$ estimations, the overall $V_{bg}$ dependence is consistent. We infer that the discrepancy could have been caused by a non-sinusoidal current-phase relationship in the SGS junction [22] and/or incomplete filtering of the environmental noise.

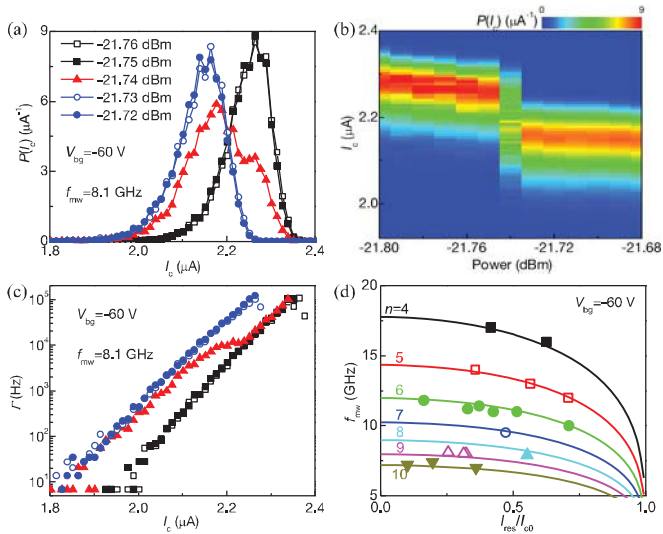

FIG. 4. (color online) (a) $P(I_c)$ under microwave irradiation of varying power. The microwave frequency was $f_{mw} = 8.1$ GHz, and $T = 50$ mK. (b) Color-coded density plot of $P(I_c)$ with varying microwave power. (c) $\Gamma(I)$ plot corresponding to the data in (a). (d) Applied microwave frequency $f_{mw}$ versus normalized resonant bias current $I_{res}/I_{c0}$ (symbols). Solid lines are best fits to the formula for the multi-photon absorption transition (see text) with the corresponding number of photons ($n = 4 - 10$) from top to bottom.

One-dimensional nanostructured proximity Josephson junctions normally support low junction critical currents (for instance, ~100 nA for nanowires [4, 5] and ~10 nA for carbon nanotues [6, 7]). The corresponding weak junction coupling energies, $E_{J0}/k_B$ ~2 K and ~0.2 K, respectively, hamper observation of the MQT and the energy level quantization in the washboard potential well. By contrast, graphene-based proximity Josephson junctions, in association with the expandable junction width, provide both high gate-tunability and strong junction coupling strength (represented by the critical current easily reaching 1−10 $\mu$A), which make graphene-based Josephson junctions a convenient system for observing the gate-tunable quantum behavior. In our device, the junction coupling strength was made even stronger by adopting $Pb_{0.93}In_{0.07}$ superconducting electrodes rather than Al that is commonly used for realizing graphene-based Josephson junctions. Proximity Josephson junctions based on two-dimensional electron gas are also gate-tunable [20], but the gate-tunability of MQT and the energy level quantization has not been reported in the system.

We thank Dr. M.-H. Bae for providing distribution-fitting codes and helpful discussion. One of us (YJD) is grateful for useful discussion with M.-S. Choi. This work was supported by the National Research Foundation of Korea (NRF) through Acceleration Research Grants R17 2008-007-01001-0 and 2009-0083380 (for HJL), and through Grant 2011-0005148 (Basic Science Research Program, for YJD) funded by the Ministry of Education, Science, and Technology.

# Supplemental Material for: Electrically Tunable Macroscopic Quantum Tunneling in a Graphene-based Josephson Junction

Gil-Ho Lee,[1] Dongchan Jeong,[1] Jae-Hyun Choi,[1, *] Yong-Joo Doh,[2, †] and Hu-Jong Lee[1, †]

[1]*Department of Physics, Pohang University of Science and Technology, Pohang 790-784, Republic of Korea*
[2]*Department of Display and Semiconductor Physics,
Korea University Sejong Campus, Chungnam 339-700, Republic of Korea*
(Dated: September 1, 2011)

## Method

Our device was prepared by forming $Pb_{1-x}In_x$ ($x = 0.07$) superconducting electrodes on a back-gated graphene/$SiO_2$ device. Details of the device fabrication and measurement results of the ac and dc Josephson effects are published elsewhere [1]. Monolayer graphene was prepared by mechanical exfoliation of natural graphite using Scotch tape and transferred onto thermally grown 300-nm-thick silicon oxide ($SiO_2$) on the surface of a heavily electron-doped silicon substrate. With conventional electron-beam lithography and thermal evaporation, we deposited 200-nm-thick electrodes of $Pb_{0.93}In_{0.07}$ and a 10-nm-thick protecting gold layer on the graphene with a spacing ($L$) of 300 nm and a width of 6.3 $\mu$m. The $Pb_{0.93}In_{0.07}$ alloy was used for the superconducting electrodes due to its preferable surface morphology compared with pure Pb as well as its high superconducting transition temperature up to 7.0 K. The bluish part of the graphene layer, with a larger aspect ratio than the junction area in Fig. 1(a), was used to confirm that the graphene layer was a monolayer by quantum-Hall resistance [1]. The device was placed on a copper cold finger, which was thermally anchored to the mixing chamber of a dilution fridge with a base temperature of 50 mK. All electrical measurement cables were carefully filtered with two-stage resistor-capacitor (RC) low-pass filters that were thermally anchored to the mixing chamber and pi filters located at room temperature. The RC and pi filters had cut-off frequencies of 10 kHz and 10 MHz, respectively. For measurements of the $I_c$ distribution, we applied a saw-tooth-like bias current with a ramping rate of $dI/dt = 775$ $\mu$A/s, and 9,000-10,000 switching events were recorded at a current resolution of 7.8 nA for fixed temperature $T$ and backgate-voltage $V_{bg}$. A threshold voltage of $V_{th} = 5$ $\mu$V was used to determine $I_c$.

## Mean free path of graphene

Fig. S1 shows the $V_{bg}$ dependence of the normal-state conductivity $\sigma$ (blue curve) of the graphene layer of our device in the junction area denoted in red in Fig. 1(a) and the corresponding mean free path (black curve) $l_m = h\sigma(\pi/n)^{1/2}/e^2$, where $n$ is the carrier density. $l_m$ is almost independent of $V_{bg}$ except for the vicinity of the charge-neutral Dirac point $V_{DP} = -2.5$ V. Near $V_{DP}$, $|V_{bg} - V_{DP}| < 5$ V, $l_m$ diverges unphysically due to the invalid assumption of the vanishing carrier density and density of states (DOS) [2] at $V_{DP}$. The range of $V_{bg}$ producing the unphysical values of $l_m$ corresponds to the carrier density of $|n| < 0.4 \times 10^{12}$ cm$^{-2}$, which is consistent with the result of a previous report [3] concerning the extra carriers induced by electron-hole fluctuations in graphene on $SiO_2$. To remedy the anomalous behavior of $l_m$ near $V_{DP}$, we simply add a finite Gaussian broadening to DOS as in Eq. (5) of C. Józsa *et al.* (Ref. 2). An energy broadening of 75 meV corresponds to our estimation of carrier density variation. Corrected mean free path $l_m^*$ (red line) is about 22 nm, which is much shorter than the junction spacing of $L = 300$ nm.

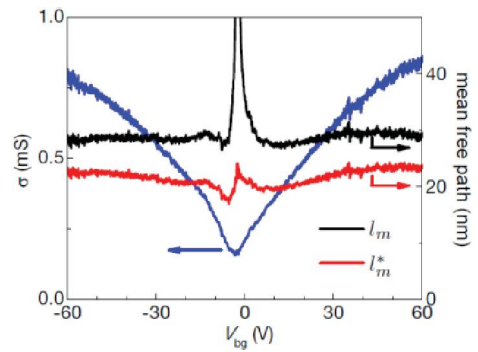

FIG. S1. (color online) Normal-state conductivity (blue) of the graphene layer and the corresponding mean free path.

## RCSJ model fit

The hysteretic current-voltage ($I - V$) curve in a Josephson junction can be understood phenomenologically by the resistively and capacitively shunted junction (RCSJ) model [4]. The time dependence of the Josephson phase is given by equating the bias current $I$ to the sum of the currents from three different parallel channels as $I = I_{c0}\sin\phi + V/R_N + CdV/dt$. Combining with the ac Josephson relation of $2eV = \hbar(d\phi/dt)$, one obtains a second-order differential equation

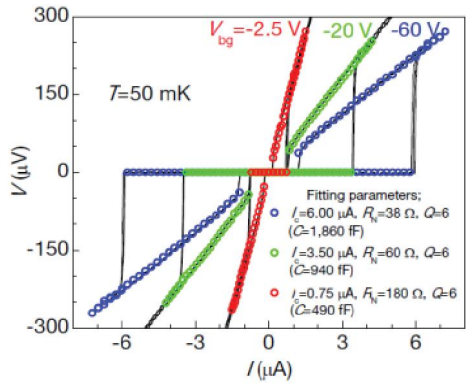

FIG. S2. (color online) $I-V$ curves (black lines) for $V_{bg} = -60$, -20, -2.5 V in comparison with the calculation results (symbols) using the RCSJ model.

$$d^2\phi/d\tau^2 + Q^{-1}d\phi/d\tau + \sin\phi = I/I_{c0}, \quad (1)$$

with a dimensionless time variable $\tau = \omega_{p0}t$, the plasma frequency $\omega_{p0} = (2eI_{c0}/\hbar C)^{1/2}$, and the quality factor $Q = \omega_p R_N C$. Figure S2 shows $I-V$ curves obtained numerically based on Eq. (1) by fitting to the experimental results at $V_{bg} = -60$, -20, -2.5 V in Fig. 1(b) with the parameter values given in the panel.

### Theoretical fit to the PD and TA model

The temperature-dependent behavior of the switching-current ($I_c$) distribution for $T > 1.80$ K represents a typical switching behavior of the phase diffusion (PD) process [5], in which the phase particle thermally activated to escape from a Josephson-junction washboard potential well is repeatedly retrapped in the neighboring potential well as a result of a strong dissipation during the escape. Since a phase particle that escapes from a potential minimum has a finite probability ($P_{RT}$) to be retrapped in the next potential minimum, the switching rate ($\Gamma_s$) in the PD regime is suppressed by the successive retrapping processes. For this PD process, $\Gamma_s$ is expressed as [5] $\Gamma_s = \Gamma_{TA}(1 - P_{RT}^{-1})\ln(1 - P_{RT})$, where $\Gamma_{TA}$ stands for a thermally activated escape rate and $P_{RT}$ is the retrapping probability obtained from integration of the retrapping rate [6, 7] $\Gamma_{RT} = \omega_p[(I - I_{r0})/I_{c0}](E_{J0}/2\pi k_B T)^{1/2}\exp(-\Delta U_{RT}/k_B T)$, with $\Delta U_{RT}(I) = (E_{J0}Q_0^2/2)[(I - I_{r0})/I_{c0}]^2$. Here, $I_{r0}$ is the fluctuation-free retrapping current, $\Delta U_{RT}$ is the retrapping barrier, and $Q_0 = 4I_{c0}/\pi I_{r0}$. Since $Q_0$ is in the exponent of $\Gamma_{RT}$, the parameters $I_{c0}$, $I_{r0}$, and $T$ dominate the fitting.

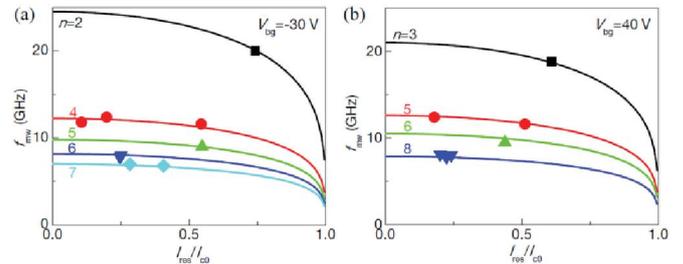

FIG. S3. (color online) Multi-photon-absorption data for (a) $V_{bg} = -30$ V and (b) $V_{bg} = 40$ V.

### Multi-photon absorption for different values of $V_{bg}$

The microwave spectroscopy data for different $V_{bg} = -30$ and 40 V are shown in Fig. S3. The resonant-condition curves are obtained for $\nu_{p0} = 49.05 \pm 0.55$ ($62.98 \pm 0.96$) GHz and $I_{c0} = 5.66 \pm 0.17$ ($6.96 \pm 0.54$) $\mu$A for $V_{bg} = -30$ (40) V. The number ($n$) of photons, which are absorbed for the transition, ranges from 2 to 8.

### Details of the fitting of $P(I_c)$ and $\Gamma(I_c)$ curves in respective escape regimes

Figure 1(c) [Fig. 2(b)] of the main text shows the fitting for the observed switching current distribution $P(I_c)$ [the escaping rate $\Gamma(I_c)$] at $V_{bg} = -60$ V in the whole temperature range used in this study. Since the horizontal scale of the switching current is highly squeezed the quality of the fits is hard to be examined. We thus reillustrate, in Figs. S4(a) and S4(b), fittings for the representative data of $P(I_c)$ and $\Gamma(I_c)$, respectively, in the separate regimes of the phase-particle escape ($T = 2.20$ K, 1.00 K, and 0.05 K) with the scale of the horizontal axes expanded sufficiently. The quality of the fitting is more conveniently resolved in this fashion. Both of fittings exhibit excellent model fits in the respective regimes of PD, TA, and MQT. The best fit parameters employed are $I_{c0}$ and $C$ for the MQT regime; $I_{c0}$, $T_{esc}$, and $C$ for the TA regime; $I_{c0}$, $I_{r0}$, $T_{esc}$, and $C$ for the PD regime. As mentioned in the main text the best-fit value of $C$ is 11.5 fF for $V_{bg} = -60$ V. The behavior of $T_{esc}$ in the TA and PD regimes is shown in the inset of Fig. 2(a) of the main text. Figure S4(c) shows the variation of the best-fit values of $I_{c0}$ and $I_{r0}$ as a function of the bath temperature $T$, along with the corresponding measured mean values, $<I_c>$ and $<I_r>$, respectively. The fluctuation-free values of the switching current [retrapping current] turns out to be slightly larger [smaller] than the corresponding mean values.

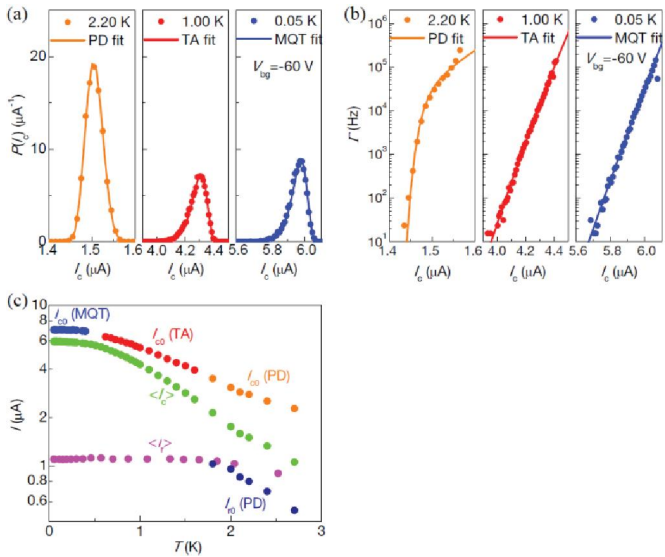

FIG. S4. (color online) Details of fitting to the MQT, TA, and PD models of (a) the measured distribution $P(I_c)$ and (b) the corresponding escaping rate $\Gamma(I_c)$. (c) Best-fit values of $I_{c0}$ and $I_{r0}$ in different escape regimes of the phase particle, along with the measured mean values of $<I_c>$ and $<I_r>$, respectively.

### Effective Capacitance

In the main text, we propose the diffusive nature of carriers in graphene as the origin of the effective capacitance, replacing the charge relaxation time $R_N C$ by the diffusion time of Andreev pairs, $\hbar/E_{Th}$ [8]. To confirm the validity of this proposition we examine the linearity between $C$ and the inverse of the normal-state resistance $1/R_N$ in the relation of $C_{eff} = \hbar/(R_N E_{Th})$ by plotting $C_{MQT}$, obtained from fitting of the switching current distribution in the MQT regime, as a function of $1/R_N$ [see Fig. S5]. We choose $C_{MQT}$ for the plot because $C$ can be determined more precisely from the switching current distribution in the MQT regime than in the TA regime as $\Gamma_{MQT}$ has an exponential dependence on $C$ whereas $\Gamma_{TA}$ does not.

A clear linear relation is obtained between $C_{MQT}$ and $1/R_N$ with the best-fit solid line, which is consistent with our proposition about the origin of the effective capacitance. But the slope, ∼0.6 ps, turns out to be smaller by an order of magnitude than the value $\hbar/E_{Th} = 8$ ps estimated with the Thouless energy $E_{Th} = 80$ $\mu$eV. This discrepancy was possibly caused by the rather crude assumption of equating $R_N C$ and $\hbar/E_{Th}$, which did not take into account any details of the pair dynamics such as multiple Andreev reflections [8]. A part of the discrepancy may also have been introduced as the washboard potential barrier $\Delta U$ was modified by the non-sinusoidal current-phase relation [9, 10].

Our superconductor-graphene-superconductor (SGS) junction was in a long-junction limit as the width of the junction, $W$=6.3 $\mu$m, was larger than the Josephson penetration depth, $\lambda_J = (\Phi_0/2\pi\mu_0 L j_c)^{1/2}$ ∼1 $\mu$m, where $\Phi_0$ is the magnetic flux quantum, $\mu_0$ is the vacuum permeability, and $j_c = I_c/(Wt)$ is the critical current density (with $t$ as the thickness of the graphene layer). In a long-junction limit, $\Delta U$ was also possibly reduced by the spatial nonuniformity of the phase difference across the junction, $\Delta\phi$, along the junction width [11–13]. Then, the capacitance $C$ estimated from the distribution fitting in the MQT regime is expected to be reduced as the square of $\Delta U$. The elevation of $T_{esc}$ above the bath temperature in the distribution fitting in the TA regime [inset of Fig. 2(a)] seems to support that the effective $\Delta U$ is indeed reduced significantly (by about 30%).

### Energy level quantization

A theoretical analysis [14] shows that more microwave power is required to excite a phase particle as the junction damping increases. As our SGS Josephson junction has a $Q$ (∼5) much lower than conventional tunneling-type Josephson junctions ($Q \geq 100$), the junction has a high energy-dissipation rate, requiring higher microwave power to excite a phase particle than in conventional high-$Q$ tunneling-type Josephson junctions. Moreover, a large number of photons involved in microwave measurements ($n$ = 2-10) leads to the occurrence of the energy level quantization (ELQ) only for high microwave power.

To confirm that the ELQ features we observed arose from the transition from the ground state to an excited state, spectroscopic data with a wider range of microwave power than in Fig. 4(a) is shown in Fig. S6(a). The initial state of each cycle of $I_c$ measurements was reset to the ground state for a time duration sufficiently longer (∼10 ms) than the dissipation time constant ($\tau_{diss} = R_N C_{eff}$ ∼10 ps). Thus, the first transition of $I_c$ around -22 dBm from the lowest microwave power corresponds to the excitation from the ground state to an excited state. The ELQ features were reproducible as shown in Figs. S6(b) and (c).

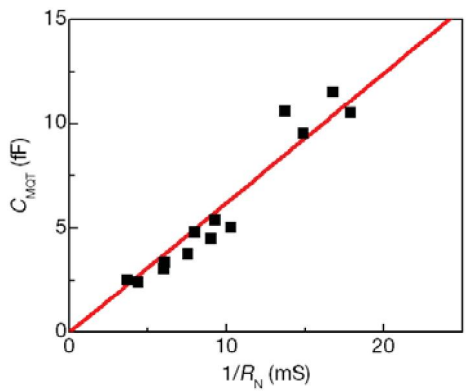

FIG. S5. (color online) A plot of $C_{MQT}$ vs $1/R_N$

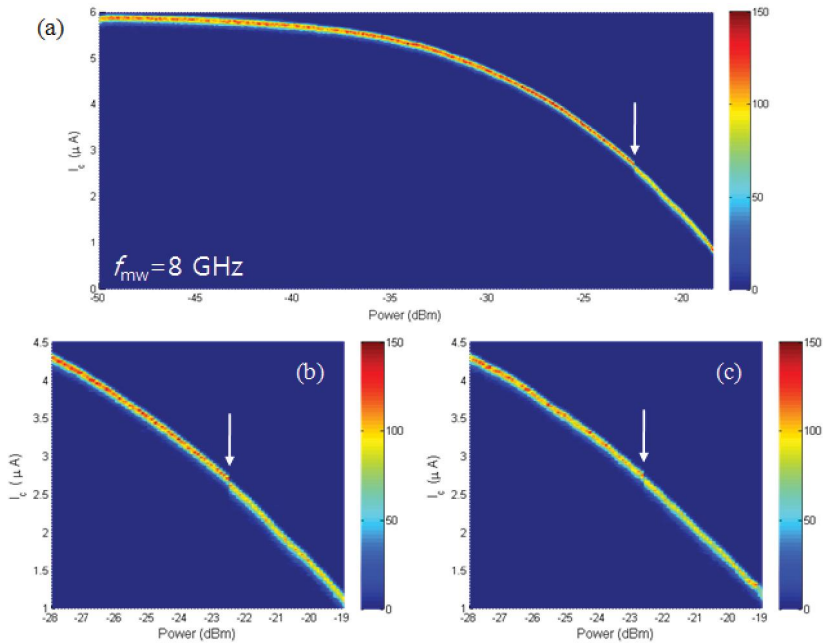

FIG. S6. (color online) (a) Color-coded plot of the number of switching events with a wide range of microwave power at $f_{mw} = 8$ GHz and $V_{bg}$=-60 V. White arrows indicate the point of resonance with excitation of the phase particle. (b) Magnified view of (a) around the resonance point. (c) The reproducible spectroscopic data taken in a condition identical to (b) but after 24 hours.

More spectroscopic data that show ELQ features at $V_{bg}$=-60 V for various microwave frequencies $f_{mw}$ are presented in Fig. S7(a)-(d). Due to the complicated coupling of microwave to the SGS Josephson-junction device, ELQ features were obtained only at a few specific frequencies of our measurement setup where the microwave-device coupling was in an optimal condition. Examples of spectroscopic data, which do not show ELQ features, are shown in Figs. S7(e) and (f).

---

* Current address; Memory Division, Samsung Electronics, Hwasung 445-701, Republic of Korea
† To whom correspondence should be addressed. E-mail: yjdoh@korea.ac.kr (Y.-J. Doh) and hjlee@postech.ac.kr (H.-J. Lee)

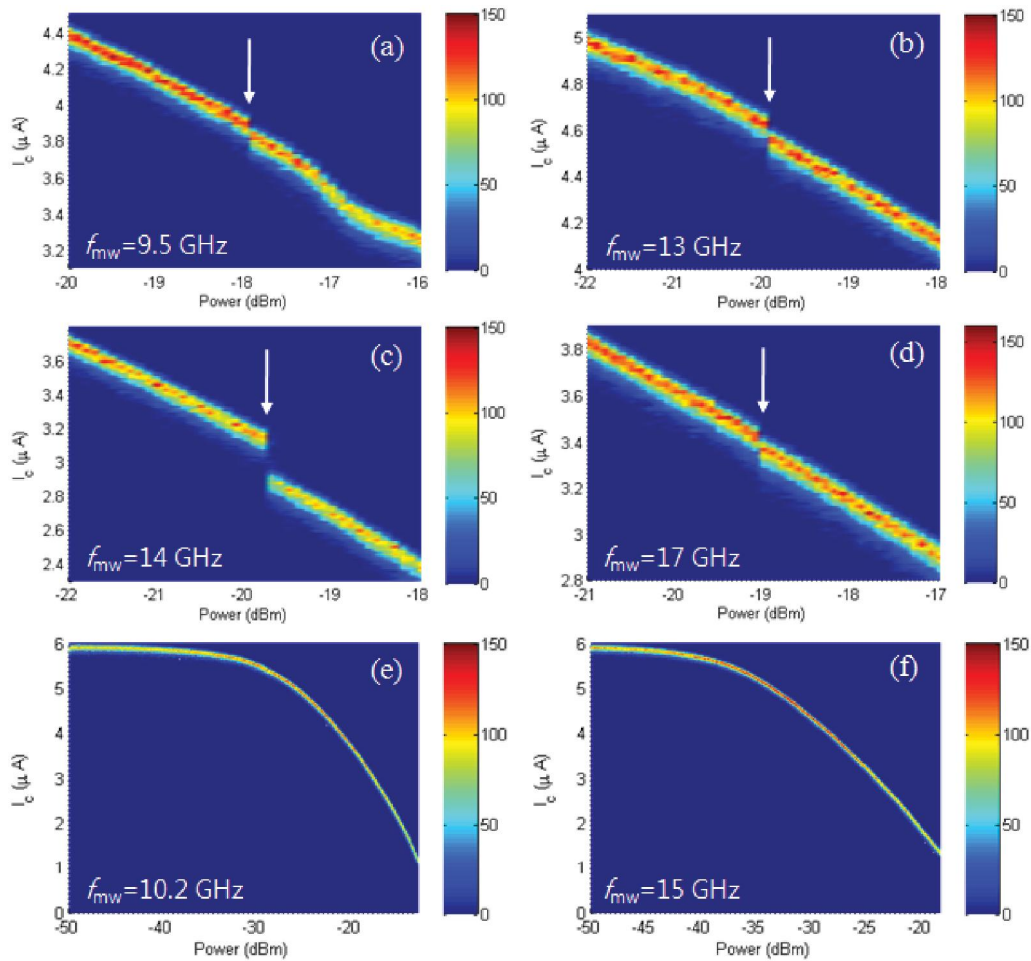

FIG. S7. (color online) (a)-(d) Spectroscopic data at $V_{bg}$ =-60 V for various microwave frequencies. White arrows indicate points of the resonance. (e) and (f) Spectroscopic data at $V_{bg}$ =-60 V, where ELQ features are not observed.